\documentclass[aps,floats,twocolumn,epsf,prl,showpacs]{revtex4-1}
\usepackage[USenglish]{babel}
\usepackage{amsmath,amssymb,amsfonts}
\usepackage{epsfig}
\usepackage{graphicx}
\usepackage{color}

\usepackage[colorlinks=true,linkcolor=blue,citecolor=red]{hyperref}

\newcommand{\eqsplit}[2]
{
  \begin{equation}
    \begin{split}
      #1
    \end{split}
    \label{#2}
  \end{equation}
}

\newcount\colveccount
\newcommand*\colvec[1]{
  \global\colveccount#1
  \begin{pmatrix}
    \colvecnext
  }
  \def\colvecnext#1{
    #1
    \global\advance\colveccount-1
    \ifnum\colveccount>0
    \\
    \expandafter\colvecnext
    \else
  \end{pmatrix}
  \fi
}

\newtoks\rowvectoks
\newcommand{\rowvec}[2]{%
  \rowvectoks={#2}\count255=#1\relax
  \advance\count255 by -1
  \rowvecnexta}
\newcommand{\rowvecnexta}{%
  \ifnum\count255>0
  \expandafter\rowvecnextb
  \else
  \begin{pmatrix}\the\rowvectoks\end{pmatrix}
  \fi}
\newcommand\rowvecnextb[1]{%
  \rowvectoks=\expandafter{\the\rowvectoks&#1}%
  \advance\count255 by -1
  \rowvecnexta
}

\def\bcen{\begin{center}}
\def\ecen{\end{center}}

\def\HH{{\cal H}}

\def\eg{\mbox{\it e.g.\ }}  \def\ie{\mbox{\it i.e.\ }}
\def\=={\equiv}

\def\qed{\raise1pt\hbox{\vrule height5pt width5pt depth0pt}}

\def\cG0{{\cal G}_0}
\def\cG{{\cal G}}

\def\spinup{\uparrow} \def\spindown{\downarrow} 
\def\up{\uparrow}  \def\dw{\downarrow}
\def\bra{\langle} \def\ket{\rangle}
\def\ka{{\bf k}}

\def\ie{\hbox{\it i.e.\ }} \def\eg{\mbox{\it e.g.\ }}
\def\ie{\mbox{\it i.e.\ }} \def\=={\equiv}

 \def\ep0{\epsilon_{p}} \def\ed0{\epsilon_{d}}

\def\bhzrowvec{\rowvec{4}{c_{1\ka\spinup}}{c_{2\ka\spinup}}{c_{1\ka\spindown}}{c_{2\ka\spindown}}}

\begin{document}

\title{First-order character and observable signatures of topological
  quantum phase transitions} 
\author{A.~Amaricci$^\mathrm{1}$}
\author{J.~C.~Budich$^\mathrm{2,3}$}
\author{M.~Capone$^\mathrm{1}$}
\author{B.~Trauzettel$^\mathrm{4}$}
\author{G.~Sangiovanni$^\mathrm{4}$}

\affiliation{$^\mathrm{1}$Democritos National Simulation Center,
Consiglio Nazionale delle Ricerche,
Istituto Officina dei Materiali (IOM) and
Scuola Internazionale Superiore di Studi Avanzati (SISSA),
Via Bonomea 265, 34136 Trieste, Italy}
\affiliation{$^\mathrm{2}$Institute for Theoretical Physics,
University of Innsbruck, 6020 Innsbruck, Austria}
\affiliation{$^\mathrm{3}$ Institute for Quantum Optics and Quantum
  Information, Austrian Academy of Sciences, 6020 Innsbruck, Austria} 
\affiliation{$^\mathrm{4}$Institut f\"ur Theoretische Physik und
  Astrophysik, 
Universit\"at W\"urzburg, Am Hubland, D-97074 W\"urzburg, Germany}

\date{ \today }

\begin{abstract}
Topological quantum phase transitions are characterized by changes in
global topological invariants. These invariants classify many-body
systems beyond the conventional paradigm of local order parameters
describing  spontaneous symmetry breaking. For non-interacting
electrons, it is well understood that such transitions are continuous
and always accompanied by a gap-closing in the energy spectrum, given
that the symmetries protecting the topological phase are
maintained. Here, we demonstrate that a sufficiently strong 
electron-electron interaction can fundamentally change the situation:
we discover a topological quantum phase transition of first-order
character in the genuine thermodynamic sense, that occurs without a gap
closing. Our theoretical study reveals the existence of a quantum
critical endpoint associated with an orbital instability  on the
transition line between a 2D topological insulator and a trivial band
insulator. Remarkably, this phenomenon entails unambiguous signatures
related to the orbital occupations that can be detected
experimentally. 
\end{abstract}


\maketitle

\emph{Introduction} -- The advent of topological insulators
\cite{HasanKane,MooreReview,XLReview} has given
a surprising twist to the venerable topic of bandstructure physics:
Triggered by the theoretical prediction
\cite{KaneMele2005a,KaneMele2005b,BHZ2006} and experimental
observation \cite{koenig2007} of the quantum spin Hall effect,
topological aspects of band insulators and 
superconductors have become one of the most active research fields in
physics, culminating in a new periodic table
\cite{Schnyder2008,Kitaev2009,Ryu2010} that exhaustively lists all
topologically distinct bandstructures in the ten Altland-Zirnbauer
symmetry classes \cite{Altland1997}. 
For the integer quantum Hall effect, the archetype of a topological state,
the discovery of the striking quantization of a
natural response quantity, namely the Hall conductance
\cite{Klitzing1980}, preceded its theoretical explanation in terms of topology \cite{Laughlin1981,TKNN1982}.
In contrast, most of the topological invariants of the periodic table are not directly related to bulk observables.
The identification of experimentally detectable signs of topological states and of phase transitions between them is hence an open challenge.

\begin{figure*}[t]
  \includegraphics[width=1\linewidth]{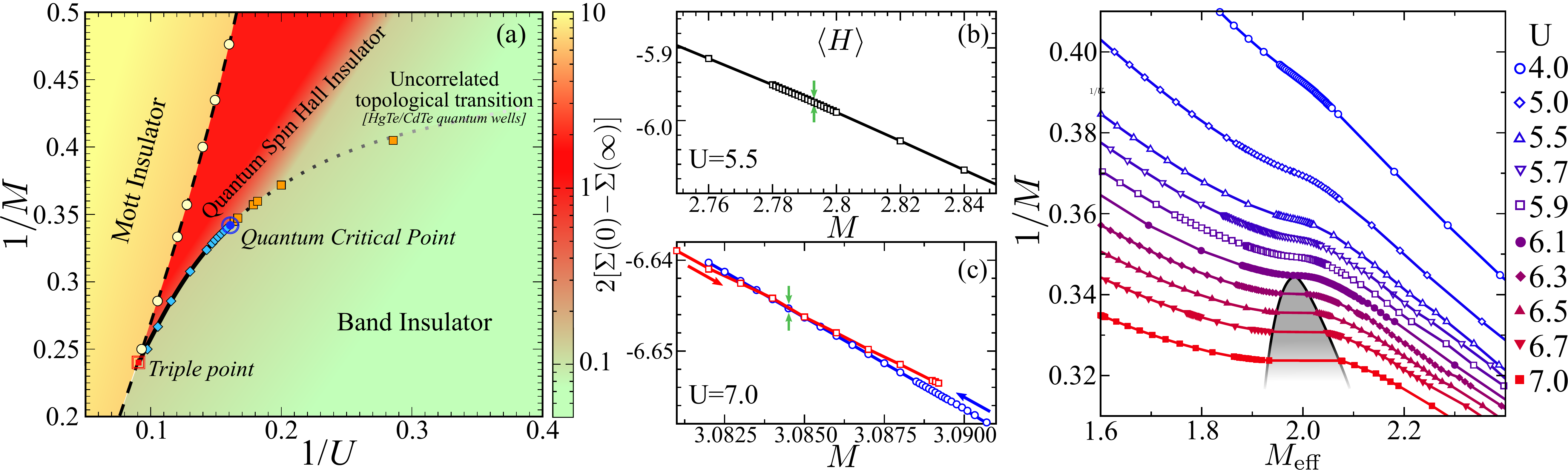}
  \caption{
  (Color online) 
    (a) $T=0$ phase diagram in the $1/U\, vs\, 1/M$ plane for $J=U/4$ and $\lambda =0.3$.
    Besides delimiting the different phases -- Mott, Topological and Band Insulator (MI, QSHI and BI, respectively, in the main text) -- the color quantifies the many-body character, as measured by the value of $\Xi\!=\!2|{\Sigma}(0)\!-\!{\Sigma}(\infty)|$.
    The orange squares and the dotted line mark the continuous BI-QSHI transition for small $U$. The blue diamonds and the thick solid line mark the first-order transition between the same phases for large $U$. The two lines are connected by a quantum critical point. White circles and a dashed line denote the boundary of the MI.
    (b) and (c) Total energy $\bra \HH\ket$ as a function
    of $M$, for two values of $U$. Vertical arrows mark the transition. 
    The red and blue curves in (c) denote the solutions coming from the QSHI and from the BI, respectively. The branches with higher energy correspond to metastable solutions that can be followed beyond the transition point and give rise to hysteresis.
    (d) Effective band splitting parameter $M_\mathrm{eff}$ near the quantum critical
    point showing the critical behavior of the transition.
  }
  \label{fig:phasediagram}
\end{figure*}

In conventional Landau-Ginzburg theory, different states of matter are classified by spontaneous symmetry breaking, associated with local order parameters.
Phase transitions of this kind are generically of first-order, i.e., the order parameter and other relevant observables display discontinuities at the phase boundary, providing us with a natural way of detecting them \cite{SenthilS2004}. 
On the contrary, topological phase transitions elude this paradigm as they are associated to a change in global invariants. 
Since in the non-interacting case, the bandstructure evolves continuously as a function of the control parameters, topological transitions are always continuous and accompanied by a closure of the energy gap, given that the protecting symmetries are preserved and that the Hamiltonian is short-ranged.
The inclusion of strong electronic correlations may however lead to new scenarios\cite{raghuPRL100,pesinNaturePhys6,wrayNatPhys7,xiaoNatComm2,rueeggPRL108,MartinFakherReview,kargarianPRL110,dengPRL111,herbutPRL113}.
In this regard, a natural and still open question is whether correlation-induced topological transitions can exhibit novel experimental signatures, compared to their non-interacting analogs.

Here, we report for the first time the occurrence of a {\emph{first-order}} topological quantum phase transition in a microscopic model. 
In a sense, strong correlations superimpose to the topological transition a conventional critical behavior, even though the transition still does not correspond to any long-range order in the ordinary thermodynamic sense.
This allows us to identify natural experimental fingerprints of the topological transition, in addition to the global invariants.
Specifically, by solving a two-orbital Hubbard model with Dynamical
Mean-Field Theory (DMFT), we show that a transition between a band insulator (BI) and a quantum spin Hall insulator (QSHI) takes
place without the continuous closing of the band gap in the electronic
spectrum if the Coulomb repulsion is large enough. In this
strongly-correlated regime, the interaction-induced QSHI is thus in
proximity to a first-order transition line and to a quantum critical endpoint associated with a correlation-driven criticality in the orbital 
channel.
A direct, yet crucial, implication of this result is the 
identification of a critical behavior in measurable quantities
related to the orbital polarization.

\emph{Model} -- We study a minimal two band Fermi-Hubbard model with both inter- and
intra-band interactions on a two-dimensional square lattice
exhibiting the quantum spin Hall effect \cite{budichPRB87}. The model
Hamiltonian reads as
\eqsplit{
&\HH = \sum_\ka \psi_\ka^\dag \, \hat{{\bf H}}_0(\ka)\,
\psi_\ka  + \sum_{\mathrm{i}} \HH_{\mathrm{int}}({\mathrm{i}}),\cr
&\HH_{\mathrm{int}}({\mathrm{i}}) = (U-J)
\frac{N_{\mathrm{i}}(N_{\mathrm{i}}-1)}{2} - J \left(
  \frac{N^2_{\mathrm{i}}}{4} + S_{z{\mathrm{i}}}^2 - 2
  T_{z{\mathrm{i}}}^2 \right)\,  
}{eq:BHZHubbard} 
where $\psi_\ka = \bhzrowvec$ and the operator $c^\dag_{\ka \alpha\sigma}$
($c_{\ka \alpha\sigma}$) creates (annihilates) an electron in orbital
$\alpha=1,2$ with momentum $\ka$ and spin $\sigma$.
$N_{\mathrm{i}}$ is the number operator for the electrons on site
${\mathrm{i}}$, summed over $\alpha$ and $\sigma$. 
$S_{z{\mathrm{i}}}$ and $T_{z{\mathrm{i}}}$ are the $z$-components of
the total spin $S_{\mathrm{i}}$ and orbital pseudo-spin $T_{\mathrm{i}}$ operators, respectively.  
When referring to the local components of these operators, we will
drop the subscript $\mathrm{i}$. The local orbital polarization, \eg
will therefore read $T_{z}\!=\!( n_{1\up} + n_{1\dw} - n_{2\up}
-n_{2\dw}  )/2 = (n_1 - n_2)/2$. 

The non-interacting part $\hat{{\bf H}}_0$ of the Hamiltonian
(\ref{eq:BHZHubbard}) conserves $S_z$, \ie it has a $U(1)$ spin
rotation symmetry and, hence, is block diagonal in spin:
$\hat{{\bf H}}_0(\ka)=\mathrm{Diag}[{\hat{h}(\ka),\hat{h}^*(-\ka)}]$.
The relation between the two blocks is imposed by time reversal
symmetry (TRS).
Denoting the orbital pseudo-spin by $\tau$, the individual blocks have
the form $\hat{h}(\ka)=d^i_\ka \tau_i$, with 
$d^x_\ka=\lambda\sin(k_x),~d^y_\ka= \lambda\sin(k_y),~
d^z_\ka=M-\cos(k_x)-\cos(k_y)$, describing a system of two bands of
width $W=4$ in our units of energy, coupled with an inter-orbital hybridization $\lambda$
and separated by an energy splitting $2M$.
We set $\lambda = 0.3$ and we have checked that
our results are qualitatively robust against this choice.  
We consider an average density of two electrons per site
(half-filling), which corresponds to setting the chemical potential
$\mu$ at the center of the gap. 

In the absence of interaction 
and with the interpretation of
$\alpha=1,2$ as orbitals in a
semiconductor quantum well, 
a similar model has been originally introduced \cite{BHZ2006} for
HgTe-CdTe heterostructures  where the quantum spin Hall effect has
first been experimentally observed \cite{koenig2007}. 
In our model the two orbitals should be regarded as localized
electron wave functions of strongly correlated systems or as lowest
Wannier functions in a deep optical lattice with fermionic atoms.
The second term $\HH_\mathrm{int}$ in (\ref{eq:BHZHubbard}) describes
the screened local Coulomb repulsion experienced by the fermions\cite{georgesHundRev}. This term 
includes both {\it inter}- and {\it intra}-orbital
interaction terms as well as the Hund's coupling $J$, in order to
account for the natural tendency of the interacting electrons to
maximize the total spin.

A non-perturbative solution of our interacting model
(\ref{eq:BHZHubbard}) in the thermodynamic limit is attainable by
means of DMFT \cite{georgesRMP}.
This approximation goes beyond the static mean-field (Hartree-Fock)
level because it treats the self-energy, which describes all
interaction effects entering the single-particle Greens function, as a
frequency dependent, though purely local, quantity.  
The orbital structure, crucial to describe many-body effects on
topology \cite{budichPRB86}, is also fully captured within
DMFT. Here, the self-energy in the orbital pseudo-spin space has the
form $\hat{\Sigma}(\omega)=\Sigma(\omega) \tau_z$, where
$\Sigma(\omega)$ is a scalar complex function. 
The real part of $\hat{\Sigma}$ at $\omega\!=\!0$ normalizes the
energy splitting between the two orbitals: 
$2M_{\mathrm{eff}}=2M + \text{Tr}[\tau_z \hat {\Sigma}(\omega\!=\!0)]$.


\emph{Results} --
The zero temperature paramagnetic phase diagram of our model (\ref{eq:BHZHubbard}) is shown in
Fig. \ref{fig:phasediagram}(a) in the $1/M$ vs $1/U$ plane. In each point of this diagram we use a color code proportional to $\Xi\!=\!2|{\Sigma}(0)\!-\!{\Sigma}(\infty)|$.
This quantity measures the deviation of the DMFT self-energy at low frequency from the high-frequency values, which essentially corresponds to the static Hartree-Fock mean-field. In other words $\Xi$ quantifies the strength of quantum fluctuations beyond static mean-field or the degree of correlation. 

At $U\!=\!0$ we recover the conventional  topological transition where the gap closes at $M=2$, separating the trivial BI from the QSHI. When the interactions $U$ and $J$ are switched on, the two phases undergo a different evolution. The BI, with two electrons in the lower-lying orbital, is essentially unaffected by the interactions, while the QSHI, in which the two orbitals are more equally occupied, is much more exposed to correlation effects.
The interactions favor equal population of the orbitals. Therefore the self-energy correction to $2M$ reduces the energy separation between them\cite{parraghPRB88} and  a larger value of $M$ is required to turn the QSHI into a trivial BI.
The QSHI-BI transition line in the phase diagram is defined as the locus where the effect of $U$ is balanced by an increase of $M$ which is given by the condition $M_{\mathrm{eff}}\!=\!2$. Crossing this line, the topological $\mathbb Z_2$-invariant, evaluated
from the single-particle Green's function \cite{WangInversion,WangZhang2012}, changes from $0$ in the BI, to $1$ in the QSHI. 

However, the nature of the transition dramatically changes from weak to strong interactions. In the small-$U$ regime the transition remains continuous (dotted line) as in the $U=0$ limit. For larger interactions the evolution between BI and QSHI turns into a discontinuous first-order transition (thick solid line). A quantum critical point (QCP)  at ($1/U_c$,$1/M_c$) separates the two regimes.
The change in the character of the transition arises  from a non-trivial competition between the single-particle term controlled by $M$ and the interaction proportional to $U$ that we can visualize  with the help of the quantity $\Xi$ defined above.
For small values of $U$, the interaction essentially behaves like a single-particle term that renormalizes the bandstructure parameters, well captured within a Hartree-Fock approximation. This means that $\Xi$ is small for both solutions (See green region on both sides of the dotted line  in Fig. \ref{fig:phasediagram}(a)). Therefore the compensation between the $M$ and $U$ terms is basically perfect and the transition maintains the properties of the non-interacting system. 
For larger values of $U$, higher-order terms of a perturbative expansion must be taken into account. In the framework of DMFT, this is reflected by a significant frequency dependence of  the self-energy and an increase of $\Xi$ for the QSHI solution\cite{budichPRB87}, which becomes rapidly more correlated as $U$ increases. On the other hand the BI remains largely unaffected by correlations (See the red region on the QSHI side of the thick solid line, to be contrasted with the BI side remaining green in Fig. \ref{fig:phasediagram}(a)). 
The frequency dependence of the self-energy implies that an increase of the $M$ term can no longer compensate exactly for the dynamical effect of the interactions, and the physical picture is no longer directly linked with the $U=0$ point.
The two ground states cannot be continuously connected and the only way to move from one to the other is a first-order jump. However, the topological characterization of the two phases remains the same as in weak coupling, and in particular the two solutions retain the values of the global topological invariants of their non-interacting counterparts.

The first-order line ends in a triple point, after which $U$ and $M$ are so large that the QSHI solution disappears in favor of a direct transition between the BI and a Mott insulator (MI), which naturally emerges when the interaction strength is larger than any other scale. This highly-correlated solution has a high-spin configuration and a very large value of $\Xi$ (Yellow in Fig. \ref{fig:phasediagram}(a)).
We have checked that, releasing the paramagnetic constraint, an antiferromagnet is stable for large $U$ but it does not spoil the critical behavior.

The first-order behavior and the associated hysteresis are further illustrated by the behavior of the internal energy $\langle\HH\rangle$ for two interactions respectively smaller (panel (b)) and larger (panel (c)) than $U_c = 6.1$. The derivative $\partial \langle \HH\rangle/\partial M \equiv 2\langle T_z\rangle$ is clearly continuous below $U_c$ and it jumps above $U_c$, but it does not vanish on either sides of the transition (See Fig. 3(a)), and it cannot be used as an order parameter. 
On the other hand, the quantity $\Delta M_\mathrm{eff}=M_\mathrm{eff}(\mathrm{BI}) - M_\mathrm{eff}(\mathrm{QSHI})$ calculated \emph{along} the topological transition line, can be viewed as an order parameter. A visual analogy with the liquid-gas transition can be obtained plotting $\Delta M_\mathrm{eff}$ as a function of $1/M$ for different values of $U$ which are the counterparts of the isotherms. The critical behavior is apparent in the corresponding plot of Fig. \ref{fig:phasediagram}(d).
The topological transition in the correlated part of the phase diagram
becomes, therefore, of first-order in the usual thermodynamic 
sense. 


 \begin{figure}[t]
     \includegraphics[width=1\linewidth]{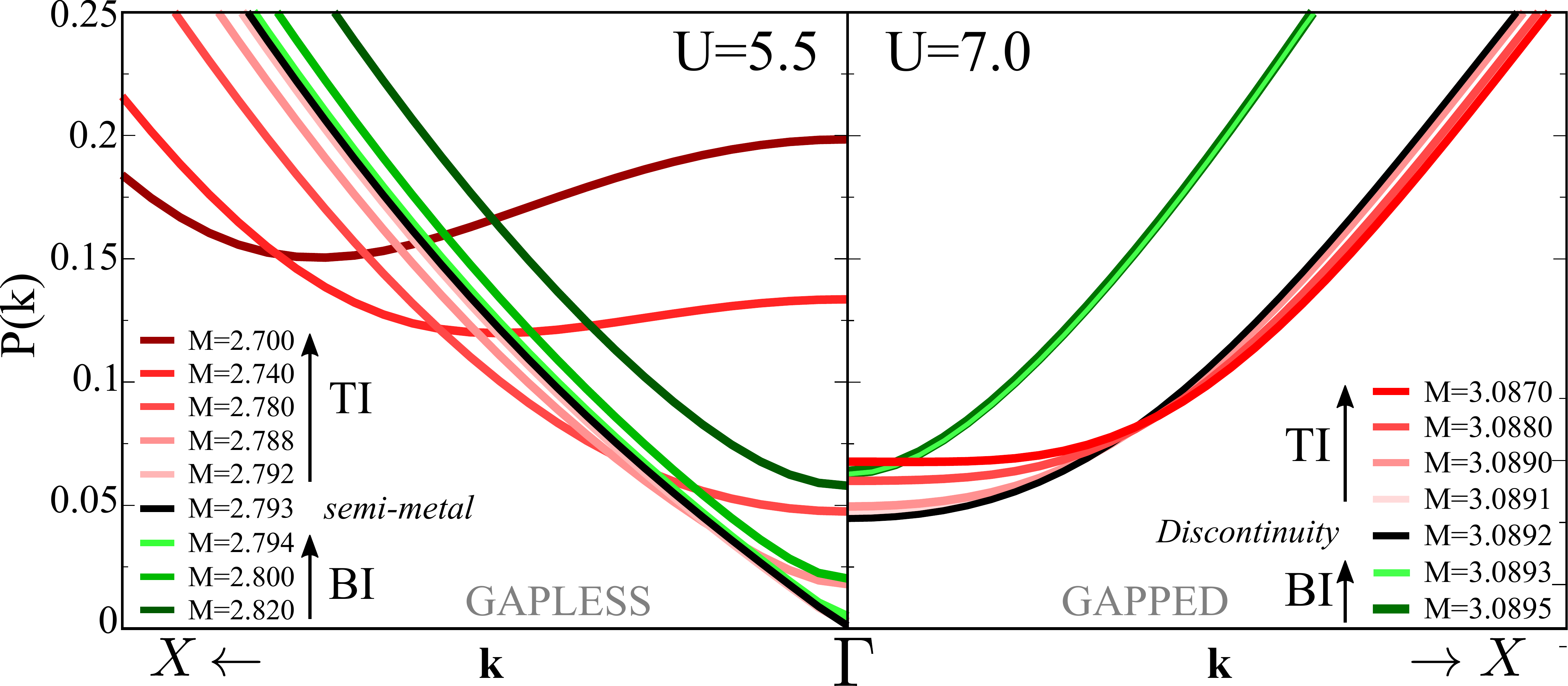}
    \caption{(Color online) Evolution of the poles $P(\ka)$ (positive part) of the
      single-particle Green's function
      near the $\Gamma$ point across the topological
      transition for $U=5.5$ (left panel) and $U=7.0$ (right panel).
      At weak-interaction (left panel) the transition occurs with a
      closure of the band-gap with the formation of a semi-metallic
      state (Dirac cone). At strong-interaction (right panel) the
      transition takes place without closure of the spectral gap. Here we follow the QSHI solution in its whole existence region.
    }
    \label{fig:GammaPoles}
  \end{figure}

 \emph{Absence of gap closing} --
 For non-interacting systems, the topological invariants are defined in terms of integrals over the compact Brillouin zone of functions of the Bloch Hamiltonian, more precisely, of the projection onto its occupied eigenstates. These functions are continuous in all bandstructure parameters as long as an energy gap is present. 
As a consequence, if the underlying symmetries of the system are maintained, the discrete-valued topological invariants cannot change without a continuous closing of the energy gap \cite{XLReview,MartinFakherReview}.
In contrast, by explicit breaking of TRS or particle number conservation \cite{rachelPRB82,ezawaSciRep3,Budich2013}, a
QSHI can be connected to a trivial band insulator without
a gap closing. In the presence of interactions, the single-particle
Green's function can acquire zeros that are also associated with
changes in the topological invariants \cite{Gurarie2011}.

Remarkably, the topological quantum phase transition line for $U>U_c$
discovered in this work does not fit into any of the above pictures. 
The topological transition 
is of first-order and is accompanied
by a discontinuity -- in the sense of a finite jump -- in the
single-particle Green's function. 
We stress that both TRS and particle number are conserved.

The absence of a gap closing is demonstrated in
Fig. \ref{fig:GammaPoles}, in which we compare the evolution of the spectral
gap around the $\Gamma$-point as a function of $M$ for a
BI-QSHI transition at $U\!=\!5.5<U_c$ and at $U\!=\!7.0>U_c$,
respectively. While a Dirac cone at the $\Gamma$-point
is present for $U=5.5$, the spectrum is gapped for $U=7.0$.
In the latter case the ground state discontinuously jumps from a gapped BI to a
gapped QSHI at the first-order transition.

  \begin{figure}
    \includegraphics[width=1\linewidth]{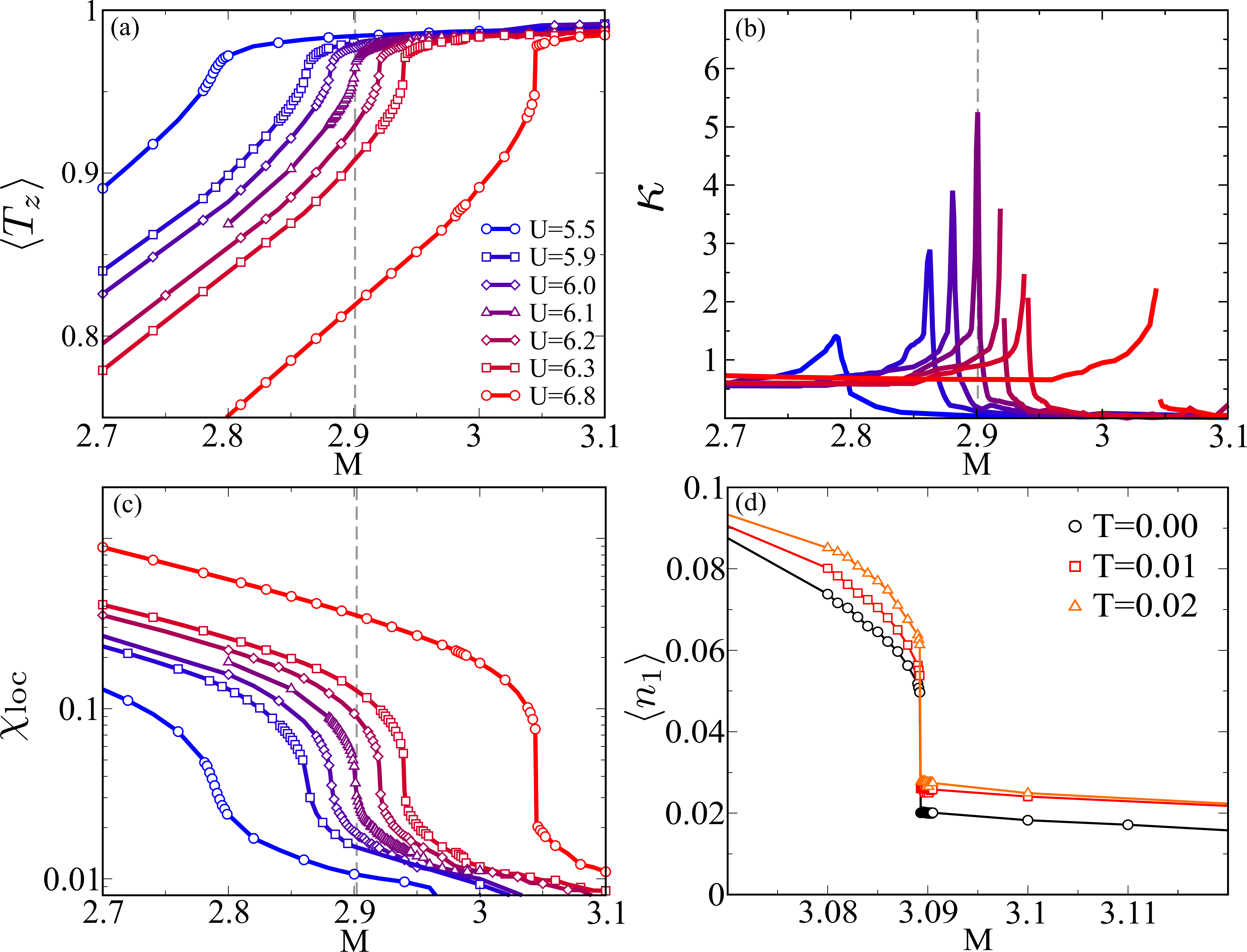}
    \caption{
      (Color online) First-order character of the topological quantum phase
      transition in the strongly correlated regime.
      (a) Local orbital polarization $\langle T_z \rangle$, 
      (b) orbital compressibility $\kappa=\partial \langle T_z 
      \rangle/ \partial M$ and (c) local spin susceptibility
      $\chi_\mathrm{loc}$ as a function $M$.
      (d) Temperature dependence of the jump in the orbital occupation
      of the first (higher-lying) orbital.
    }
    \label{fig:Exp}
  \end{figure}
The lack of the zero-gap semi-metallic state in the \emph{bulk}
electronic structure
is a
striking dissimilarity between conventional (continuous) and
correlation-driven first-order topological transitions. This
distinction is, in principle, directly measurable in photoemission
experiments \cite{WojekPRB2014}. 
Yet, the thermodynamic characterization of the first-order phase transition for $U>U_c$, enables the identification of natural experimentally detectable differences between the trivial and the non-trivial phase.

Since the QCP discovered here is associated with a critical behavior
of $M_\mathrm{eff}$, it directly influences the orbital polarization. 
 $\langle T_z \rangle$ indeed changes
from a continuous dependence on $M$, at small $U$, to a
critical one (infinite slope) at $U_c$ and eventually  displays a
first-order jump beyond the QCP (Fig.~\ref{fig:Exp}(a)). 
We can quantify this behavior by looking at the orbital compressibility 
$\kappa \!=\! \partial\langle T_z \rangle / \partial M$ (Fig.~\ref{fig:Exp}(b)).
$\kappa$ displays a maximum at the topological transition, which gets
sharper upon increasing $U$ and eventually diverges at the QCP. 
The divergence of the orbital compressibility is an ideal marker of the
thermodynamic distinction between BI and correlated QSHI.

We can identify specific
signatures of the interaction-induced topological phase transition
also in the spin sector. 
Unlike conventional band insulators, Mott systems are characterized by
the presence of (instantaneous) local magnetic moments and by a
Curie-like ($\propto 1/T$) local spin susceptibility $\chi_{\text{loc}} = \int  d\tau \langle T_\tau
S_z(\tau) S_z(0)\rangle$. 
Remarkably the characteristic critical behavior shows up also in this response
function, displayed in Fig.~\ref{fig:Exp}(c). 

At last, we demonstrate that the first-order character of the topological
transition for $U\! >\! U_c$ persists at finite temperatures. 
As reported in Fig.~\ref{fig:Exp}(d), the orbital occupation changes
slightly with $T$ but the jump remains visible. 
The discontinuity is therefore resilient to temperatures of the order of the topological gap, which is set by the hybridization $\lambda$.

\emph{Conclusions} -- We have given strong evidence that, in the presence of strong
electronic correlations, the topological phase transition from a trivial BI to a QSHI
acquires a first-order thermodynamic character.
The immediate consequence of this is the presence of a well-defined critical
behavior of simple bulk quantities such as the orbital
compressibility and the local spin susceptibility.
Finally, the analogy with liquid-gas transition suggests a
possible strategy to design and realize a correlation-induced
QSHI. Let us consider a heterostructure of a MI and a
non-inverted, \ie trivial, BI. 
Because of the proximity to the Mott region, the BI layers close to the interface
can increase their correlation strength \cite{borghiPRB81}, leading to a decrease of the effective mass term.
Near the first-order line this can eventually trigger a 
discontinuous topological transition of such few layers.

\paragraph*{Acknowledgments.} We thank F.~Assaad, M.~Fabrizio,
M.~Punk, P. Thunstr\"om and A.~Toschi for useful discussions.
A.A. and M.C. are supported by the European Union under FP7 ERC
Starting Grant n.240524 ``SUPERBAD''. 
G.S. acknowledges financial support from the DFG research units FOR1162 and FOR1346," as "G.S. acknowledges financial support by the Deutsche Forschungsgemeinschaft through FOR 1346 and FOR 1162, J.C.B. from
the ERC Synergy Grant ``UQUAM'' and B.T. from the DFG-JST research
unit ``Topotronics'', the DFG priority program SPP1666, the Helmholtz
Foundation (VITI) and the ENB graduate school on ``Topological
Insulators''.

\bibliography{localbib}

\end{document}